\documentclass[twocolumn,pra,notitlepage,superscriptaddress,reprint, amsmath,
 amssymb,aps]{revtex4-2}

\usepackage{graphics,epsfig,amsfonts,amssymb,amsmath,color}
\usepackage[normalem]{ulem}
\usepackage{bm,bbm}
\usepackage[mathcal]{euscript}
\usepackage{xcolor}
\usepackage{dsfont}
\usepackage{soul}

\usepackage{float}

\newcommand{\ket}[1]{\ensuremath{|{#1}\rangle}}

\begin{document}

\title{Localization behavior in a Hermitian and non-Hermitian Raman lattice}

\author{Entong Zhao}%
\email{ezhaoaa@connect.ust.hk}
\affiliation{%
 Department of Physics, The Hong Kong University of Science and Technology, Clear Water Bay, Kowloon, Hong Kong, China
}

\author{Yu-Jun Liu}%
\affiliation{%
 Department of Physics, The Hong Kong University of Science and Technology, Clear Water Bay, Kowloon, Hong Kong, China
}%

\author{Ka Kwan Pak}%
\affiliation{%
 Department of Physics, The Hong Kong University of Science and Technology, Clear Water Bay, Kowloon, Hong Kong, China
}%

\author{Peng Ren}%
\affiliation{Department of Physics and Astronomy, Rice University, Houston, TX, USA}

\author{Mengbo Guo}%
\affiliation{%
 Department of Physics, The Hong Kong University of Science and Technology, Clear Water Bay, Kowloon, Hong Kong, China
}%

\author{Chengdong He}%
\affiliation{Department of Physics and Astronomy, Rice University, Houston, TX, USA}

\author{Gyu-Boong Jo}%
\email{gbjo@rice.edu}
\affiliation{Department of Physics and Astronomy, Rice University, Houston, TX, USA}
\affiliation{Smalley-Curl Institute, Rice University, Houston, TX, USA}%
\affiliation{%
 Department of Physics, The Hong Kong University of Science and Technology, Clear Water Bay, Kowloon, Hong Kong, China
}

\begin{abstract}	
We propose a flexible Raman lattice system for alkaline-earth-like atoms to theoretically investigate localization behaviors in a quasi-periodic lattice with controllable non-Hermiticity. Our analysis demonstrates that critical phases and mobility edges can arise by adjusting spin-dependence of the incommensurate potentials in the Hermitian regime. With non-Hermiticity introduced by spin-selective atom loss, our calculations reveal that critical localization behaviour in this system can be suppressed by dissipation. Our work provides insights into interplay between quasi-periodicity and non-Hermitian physics, offering a new perspective on localization phenomena.

\end{abstract}

\maketitle

\section{Introduction}

In the presence of a disorder potential, quantum systems can exhibit exponential localization of wave functions induced by wave interference, a phenomenon known as Anderson localization~\cite{anderson1958absence}. Ultracold atom systems, with their exceptional purity and high controllability, enable detailed exploration of localization phenomena. Recent experiments have investigated Anderson localization in optical lattices with either random disorder~\cite{billy2008direct} or incommensurate quasi-periodic structures~\cite{roati2008anderson, luschen2018single, xiao2021observation}.

For ultracold atoms, quasi-periodic optical lattices are created by superimposing two optical lattices with incommensurate wavelengths.  In such setups, a notable transition from extended to localized states, commonly referred to as the Anderson transition, can manifest even within one-dimensional configurations~\cite{grempel1982localization, aulbach2004phase},  contrary to uncorrelated disordered systems where such transitions are expected to occur only in dimensions higher than two according to the scaling theory~\cite{thouless1974electrons,abrahams1979scaling}. 
These systems also allow for the engineering of mobility edges-boundaries that separate coexisting localized and extended states at discrete energy levels~\cite{sarma1988mobility, evers2008anderson,biddle2010predicted, luschen2018single, an2018engineering, an2021interactions, molignini2025stability}. Recent theoretical studies suggest even more complex behavior, including intermediate critical phases and hybrid phases where localized, extended, and critical states coexist in specific one-dimensional quasi-periodic models~\cite{wang2020realization, wang2022quantum, zhou2023exact, zhou2025fundamental, wang2021many}.

Recent years have seen growing interest in non-Hermitian physics~\cite{El-Ganainy2018,Ashida2021} across multiple research domains, including electrical circuits~\cite{helbig2020generalized, ezawa2019non}, optics and photonics~\cite{Lee2009,xiao2020non, zhou2018observation, weidemann2020topological}, optomechanics~\cite{Patil2022}, NV centers~\cite{zhang2021observation, yu2022experimental}, trapped ions~\cite{cao2023probing, zhang2025observation}, superconducting qubits~\cite{chen2021quantum} and ultracold atom systems~\cite{Li2019,liang2022dynamic, ren2022chiral}. Non-Hermitian systems display unique features such as exceptional points or rings~\cite{zhen2015spawning, ren2022chiral,liu2025}, novel topological invariants, restoration of the dark state~\cite{zhou2025recovering}, and the non-Hermitian skin effect~\cite{liang2022dynamic, zhao2023two, molignini2023anomalous}. The Hatano-Nelson model, a fundamental non-Hermitian topological model, combines asymmetric hopping amplitudes with an on-site random potential~\cite{hatano1996localization}. However, a comprehensive understanding of the interplay between disorder and non-Hermiticity, particularly the influences of dissipation on localization behavior, remains largely unexplored~\cite{zeng2017anderson, zhang2020non, harter2016pt}. Furthermore, experimental methods for studying this relationship between disorder and non-Hermiticity remain limited.

In this paper, we propose a flexible Raman lattice scheme~\cite{liu2013manipulating} tailored for $^{173}$Yb atoms~\cite{song2018observation,Song2019,Zhao2022} or alkaline-earth metal atoms~\cite{Liang2023} to theoretically investigate localization behaviors. The Raman lattice configuration induces quasi-periodicity using two optical lattices with incommensurate wavelengths, with the weak lattice potential akin to an incommensurate Zeeman potential. Our model shows that adjusting the frequency of the incommensurate lattice close to the $^1S_0(F=5/2)\rightarrow{}^3P_1(F=7/2)$ transition enables spin-dependent control, which can theoretically lead to the emergence of critical phases or mobility edges within the system. We further extend the Raman lattice model with incommensurate potentials into the non-Hermitian regime with dissipation induced by atom loss~\cite{zhao2023two}. Theoretical analysis reveals that dissipation exerts a significant influence on the localization behaviors within the system, providing a platform to model the intricate interplay between disorder potential and non-Hermitian physics.

The subsequent sections of this paper are structured as follows: Sec.II presents the theoretical model Hamiltonian of the flexible Raman lattice and outlines its implementation scheme. In Sec.III-VI, we theoretically analyze the impact of the spin-dependency of the incommensurate lattice and dissipation within the system. Our focus will encompass three distinct scenarios: a completely spin-dependent incommensurate lattice, a partially spin-dependent incommensurate lattice, and an incommensurate lattice coupled with dissipation. Finally, Sec.VII offers a concise summary of the theoretical findings and conclusions derived from this investigation.

\section{Model Hamiltonian}\label{Sec.II}

The Hamiltonian considered in this paper can be written as: 

\begin{equation}
	H=H_0+\sum_{j,\sigma=\uparrow,\downarrow} (\delta_{\sigma}^{j}+i\gamma_\sigma)n_{j, \sigma}
\end{equation}

\noindent where $\delta^{j}_{\sigma}=M_{z,\sigma}\cos{\left(2\pi\beta j\right)}$, with $M_{z,\sigma}\propto V_{s,\sigma}$ and $\gamma_\sigma$ denotes the incommensurate Zeeman potential strength and dissipation strength for spin $\sigma=\uparrow,\downarrow$, respectively. $n_{j, \sigma}=c^\dag_{j, \sigma}c_{j, \sigma}$ is the particle number operator, $\beta$ is an irrational number and:

\begin{equation}
	\begin{aligned}
		H_0&=-t_0\sum_{j, \sigma}\left[\left(c_{j, \uparrow}^{\dagger}c_{j+1, \uparrow}-c_{j, \downarrow}^{\dagger}c_{j+1, \downarrow}\right)+\text{H.c.}\right]\\
		&+t_{\text{so}}\sum_j\left[\left(c_{j, \uparrow}^{\dagger}c_{j+1, \downarrow}-c_{j, \uparrow}^{\dagger}c_{j-1, \downarrow}\right)+\text{H.c.}\right]
	\end{aligned}
\end{equation}

\noindent Here, $t_0$ ($t_{so}$) represents the spin-conserved (spin-flip) hopping strength between neighboring sites. The Hamiltonian renders a 1D AIII class topological insulator if replacing the incommensurate Zeeman potential with a constant uniform Zeeman potential~\cite{song2018observation, liu2013manipulating, wang2018dirac, zhang2018spin}.

\begin{figure}[htbp]
	\centering
	\includegraphics[width=1.0\columnwidth]{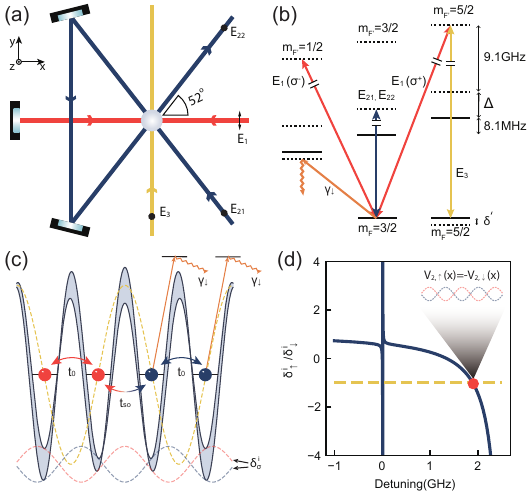}
	\caption{\textbf{Experimental implementation scheme of critical phase in optical Raman lattice with $^{173}$Yb atoms.} (a) Experimental setup of the optical Raman lattice consists of a standard 1D optical Raman lattice $E_1$, $E_3$ with another spin-dependent incommensurate lattice $E_{21}$ and $E_{22}$. (b) Schematic energy diagram with relevant transitions. (c) The incommensurate lattice induces a spin-dependent offset $V_{2,\sigma}(x)$ to the standard optical Raman lattice potential. (d) For the implementation of a perfect spin-dependent incommensurate lattice, the laser detuning should be set to approximately 1.9 GHz blue-detuned from the $^1S_0(F=5/2)\rightarrow{}^3P_1(F=7/2)$ transition.}
	\label{Fig:CP_ExperimentalSetup}
\end{figure} 

This Hamiltonian can be experimentally implemented through the configuration outlined in Fig.~\ref{Fig:CP_ExperimentalSetup}(a). To establish the conventional optical Raman lattice potential~\cite{liu2013manipulating, Wu2016,song2018observation, zhang2017two, lang2017nodal}, a standing-wave beam $E_1$ with $y$ polarization and a free-running beam $E_3$ with z polarization are employed. The single-photon detuning of these two beams can be finely tuned to approximately 11 GHz blue-detuned from the $^1S_0(F=5/2)\rightarrow{}^3P_1(F=7/2)$ transition, as depicted in Fig.~\ref{Fig:CP_ExperimentalSetup}(b). In this setup, the primary lattice, characterized by a wave vector of $k_0=2\pi/\lambda$, exhibits nearly spin-independent lattice potential $V_{1,\sigma}(x)=V_{p,\sigma}\cos^2{k_0x}$ with $V_{p,\uparrow}/V_{p,\downarrow}\sim1.01$, where $V_{p,\uparrow}/V_{p,\downarrow}$ denotes the primary lattice depth for spin up or spin down atoms. In the subsequent discussion, we will neglect the spin-dependent effects of the primary lattice potential ($V_{p,\uparrow}=V_{p,\downarrow}=V_p$). Moreover, two additional beams propagating at an angle of $\pm$52 degrees relative to the x direction are utilized to induce a spin-dependent incommensurate lattice featuring a wave vector of $k_1=k_0\cos{\theta}$, giving rise to a secondary incommensurate Zeeman potential $V_{2,\sigma}=V_{s,\sigma}\cos^2{k_1x}$ with $V_{s,\sigma}$ representing the incommensurate lattice depth, as illustrated in Fig.~\ref{Fig:CP_ExperimentalSetup}(c). The spin-dependence of this incommensurate lattice can be controlled through the detuning $\Delta$ of this lattice beam with respect to the $^1S_0(F=5/2)\rightarrow{}^3P_1(F=7/2)$ transition. When $\Delta\sim1.9
~\text{GHz}$, the incommensurate lattice becomes entirely spin-dependent with $M_{z,\uparrow}/M_{z,\downarrow}=-1$ (Fig.~\ref{Fig:CP_ExperimentalSetup}(d)). To achieve a complete spin-independent configuration, the detuning of the incommensurate lattice can be set larger than 13GHz. The Raman coupling in this setup can be denoted by $\mathcal{M}(x)=M_R\cos{k_0x}$. The introduction of dissipation into this Raman lattice setup can be realized through adding a nearly resonant loss beam with wavelength close to $^1S_0(F=5/2)\rightarrow{}^3P_1(F=7/2)$ transition~\cite{zhao2023two}.

For simplicity, all phases along the y direction in these potentials are neglected during the calculation and the Hamiltonian governing this experimental setup can be expressed as:

\begin{equation}
\begin{aligned}
H&=\left[\frac{\hbar^{2} k_{x}^{2}}{2m}+V_{1}(x)\right]\otimes\mathds{1}+\sum_{\sigma=\uparrow,\downarrow}V_{2, \sigma}(x)|\sigma\rangle\langle\sigma|\\
&+\mathcal{M}(x)\sigma_x+\frac{\delta}{2} \sigma_{z}+\sum_{\sigma=\uparrow,\downarrow}\frac{i}{2}\gamma_\sigma|\sigma\rangle\langle\sigma|
\end{aligned}
\end{equation}

\noindent Here, $\frac{\hbar^{2} k_{x}^{2}}{2m}$ is the kinetic energy term and $\delta$ denotes the two-photon detuning. The incommensurate lattice introduces a weak spin-dependent energy offset characterized by the irrational number $\beta=\cos\theta\approx0.6157$, where $\beta$ is determined by the tilted angle. In our proposed configuration, which employs $m_F$ states within the same hyperfine manifold as $\ket{\uparrow, \downarrow}$, the spin-dependence of the incommensurate lattice can be independently tuned by manipulating the detuning of the lattice beam. This independent controllability is advantageous, as it becomes challenging to simultaneously tune $\beta$ and maintain a spin-dependent incommensurate lattice when $\beta$ is set by the laser wavelength ratio. As a result, our approach significantly broadens the range of accessible $\beta$ values.

Notably, our method differs from existing schemes reported in~\cite{wang2020realization, wang2022quantum, zhou2025fundamental}. Those prior proposals were designed for alkali-metal atoms and rely on counter-propagating laser beams with wavelengths near the D1 and D2 transitions,  rendering such schemes incompatible with alkaline-earth atoms due to the lack of these transitions~\cite{He2019}. In those setups, the incommensurability parameter $\beta$ is determined by the ratio of the laser wavelengths, inherently restricting it to a narrow, specific range. Furthermore, spin dependence of the incommensurate lattice is controlled via the polarization of the laser light, which requires the spin states to be encoded in different hyperfine states. This contrasts with our configuration and previously demonstrated experimental realizations, where spin states are defined using different $m_F$ states within the same hyperfine manifold. Our scheme is particularly suited for fermionic alkaline-earth atoms such as 
$^{173}$Yb, a system already demonstrated in optical Raman lattices~\cite{song2018observation, Song2019, zhao2023two}. The fermionic nature of $^{173}$Yb, together with the potential SU(N) symmetry of alkaline-earth atoms, can lead to phenomena distinct from bosonic systems, especially in the many-body regime~\cite{molignini2025stability2, wang2021many, wang2020realization}.

\section{Localization behavior in the Hermitian regime}

Except for the localized phase, recent theoretical investigations have predicted the existence of a third fundamental phase, called critical phases, positioned between the localized and extended phases within such a spin-orbit coupled optical lattice featuring an incommensurate Zeeman potential~\cite{wang2020realization, song2018observation, zhao2023two, zhou2025fundamental}. Noteworthy features of these critical phases encompass critical spectral statistics~\cite{geisel1991new, machida1986quantum, bertrand2016anomalous}, multifractal properties of wave functions~\cite{halsey1986fractal, mirlin2006exact, dubertrand2014two}, and dynamical evolutions~\cite{hisashi1988dynamics, ketzmerick1997determines, larcher2009effects}. Moreover, recent theoretical investigations have also unveiled the possibility of a quantum phase harboring coexisting localized, extended, and critical regions within a one-dimensional quasi-periodic model~\cite{wang2022quantum}.

Therefore, we first consider an entirely spin-dependent incommensurate lattice ($M_{z,\uparrow}/M_{z,\downarrow}=-1$) in the Hermitian regime ($\gamma_\sigma=0$), and find that a critical phase manifests in this configuration. The phase of this system can be characterized through the mean fractal dimension $\bar{\eta}$, as introduced by~\cite{wang2020realization}:

\begin{equation}
	\bar{\eta}=-\frac{\lim_{L\rightarrow\infty}\ln{\left[(2L)^{-1}\sum^{2L}_{m=1}\sum^{L}_{j=1}\left(u^4_{m,j}+v^4_{m,j}\right)\right]}}{\ln{2L}}
\end{equation} 

\noindent Here, $\bar{\eta}$ is averaged over all eigenstates, which can be represented as:

\begin{equation}
	\ket{\psi_m}=\sum_{j=1}^{L}\left(u_{m,j}c^\dagger_{j,\uparrow}+v_{m,j}c^\dagger_{j,\downarrow}\right)\ket{\text{vac}}
\end{equation}

\noindent It is established that, in extended (localized) states, the value of $\bar{\eta}$ tends towards 1 (0), while within the critical phase, $0 < \bar{\eta} < 1$. 

\begin{figure}[htbp]
	\centering
	\includegraphics[width=1.0\columnwidth]{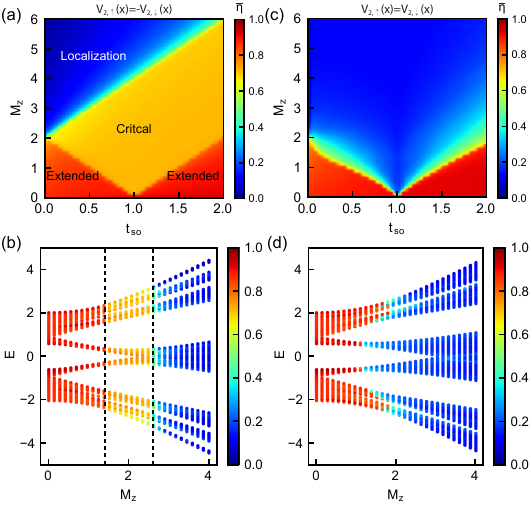}
	\caption{\textbf{Phase diagram of the optical Raman lattice system with additional incommensurate lattice.} (a) The various phases, namely extended, critical, and localization phases, are distinguished based on the mean fractal dimension $\bar{\eta}$ when $M_{z,\uparrow}/M_{z,\downarrow}=-1$. $M_z$ and $t_{so}$ are presented in units of $t_0$. (b) Fractal dimension $\eta$ and energy of individual states for distinct values of $M_z$ with $t_{so}=0.3$ and  $M_{z,\uparrow}/M_{z,\downarrow}=-1$. (c) Phase diagram of mean fractal dimension $\bar{\eta}$ as a function of $M_z$ and $t_{so}$ when $M_{z,\uparrow}/M_{z,\downarrow}=1$. Other parameters are identical to those in b. (d) Fractal dimension $\eta$ and energy of individual states for distinct values of $M_z$ with $t_{so}=0.3$ and $M_{z,\uparrow}/M_{z,\downarrow}=1$. All results are simulated with $L=1597$ and $\beta=987/1597$ in periodic boundary condition.}
	\label{Fig:CP_PhaseDiagram}
\end{figure}

In Fig.\ref{Fig:CP_PhaseDiagram}(a),  the mean fractal dimension is depicted as a function of $M_z$ and $t_{so}$, revealing a distinct region where $0 < \bar{\eta} < 1$, indicating the emergence of the critical phase. The phase boundary can be delineated by $M_z=2|t_0\pm t_{so}|$, where the symbol $+$ signifies the boundary between localization and critical phases, and $-$ signifies the boundary between extended and critical phases~\cite{wang2020realization}. Fig. \ref{Fig:CP_PhaseDiagram}(b) shows the fractal dimension for various eigenenergies and $M_z$ when $t_{so}=0.3$, illustrating that all individual states within the critical region exhibit critical behavior. However, if the incommensurate lattice is entirely spin-independent ($M_{z,\uparrow}/M_{z,\downarrow}=1)$, the critical phase region disappears, as illustrated in Fig.~\ref{Fig:CP_PhaseDiagram}(c).  Although a region where $\bar{\eta}$ deviates significantly from 1 and 0 can be identified, a further examination of the fractal dimension $\eta$ of the corresponding eigenstates as a function of $M_z$ suggests an intermediate regime characterized by the coexistence of extended and localized states rather than a critical phase (Fig.\ref{Fig:CP_PhaseDiagram}(d)).

\begin{figure}[htbp]
	\centering
	\includegraphics[width=1.0\columnwidth]{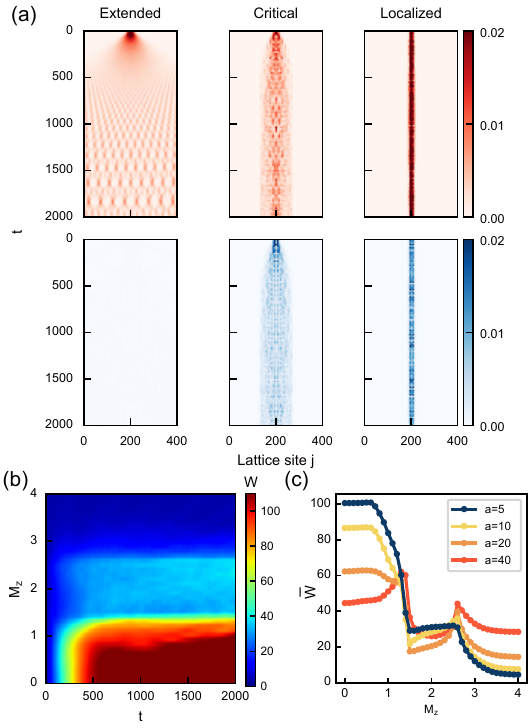}
	\caption{\textbf{Simulation of detecting critical phase with expansion dynamics.} (a) Time evolution of the wave packet for the system located in the extended phase ($M_z=0.5$), critical phase ($M_z=2.0$), and localized phase ($M_z=3.5$). (b) Expansion dynamics of the wave packet, characterized by $W$, as a function of $M_z$, with $t_{so}=0.3$ and an initial width of $a=5$. (c) Time-averaged $\bar{W}$ as a function of $M_z$ with different initial width $a$. All results are simulated with $L=400$ and $\beta=\cos{52^\circ}$ under open boundary condition.}
	\label{Fig:CP_ExperimentalDetection}
\end{figure}

In the pursuit of experimentally detecting localization behavior in an optical lattice system, one common method involves observing the expansion dynamics, as conducted in various studies \cite{billy2008direct, roati2008anderson, luschen2018single}. Fig. \ref{Fig:CP_ExperimentalDetection}(a) shows simulated time evolutions (in units of $\hbar/t_0$) of wave packets positioned in the extended, critical, and localized phase regions. These simulations are based on an initial wave packet configuration defined as $\ket{\psi_j(t=0)}=(\sqrt{\pi}a)^{-1/2}e^{-(j-j_0)^2/2a^2}\ket{\uparrow}$, where $a$ represents the half-width, and the packet is centered at site $j_0$. Notably, the outcomes reveal distinctive behaviors: in the extended phase, the atom cloud exhibits ballistic expansion, while in the localized phase, occupancy remains predominantly restricted to the initial site. In the critical phase, a quasi-localization phenomenon is observed, with the atom cloud spreading to neighboring unit cells around the initial site, without fully occupying the entire system.

To further characterize the expansion dynamics, the mean square displacement can be employed as an observable, defined as~\cite{wang2020realization}:

\begin{equation}
	W(t)=\left[\sum_{j,\sigma}(j-j_0)^2\langle n_{j,\sigma}(t)\rangle\right]^{1/2}
\end{equation}

\noindent In Fig.\ref{Fig:CP_ExperimentalDetection}(b), the evolution of $W$ over time for various $M_z$ values, with $a=5$, is presented. In the extended phase, $W$ rapidly converges to a large and stable value, while in the localized phase, $W$ remains consistently small. Conversely, in the critical phase, $W$ exhibits gradual and slow growth, distinguishing it from both the extended and localized regimes. Furthermore, the determination of phases can be facilitated by the time-averaged observable $\bar{W}=\frac{1}{N_t}\sum_{m=1}^{N_t}W(m\Delta T)$, where $N_t=100$ and $\Delta t=20$, representing the average width of the wave packet throughout the temporal evolution, as depicted in Fig. \ref{Fig:CP_ExperimentalDetection}(c). 

\section{Spin dynamics and density imbalance}

In most localization behavior studies, a crucial consideration lies in the necessity for the initial state to populate only a few lattice sites to ensure clarity in distinguishing between different phases (Fig.\ref{Fig:CP_ExperimentalDetection}(c)). While experimental setups in ultracold atom investigations enable the detection and preparation of Gaussian wavepackets as initial states through techniques like a matter wave magnifier~\cite{murthy2014matter, asteria2021quantum} or a quantum gas microscope~\cite{bakr2009quantum, sherson2010single, miranda2015site, haller2015single, edge2015imaging, omran2015microscopic, cheuk2016observation, parsons2016site, yamamoto2016ytterbium, mitra2018quantum, yamamoto2020single, kwon2022site}, this can introduce complexity and inconvenience to the experimental arrangement. To address this constraint, we also numerically simulate the evolution of spin dynamics and density imbalance across varied initial conditions, thereby offering complementary pathways to distinguish phases within the current system.

\begin{figure}[htbp]
	\centering
	\includegraphics[width=1.0\columnwidth]{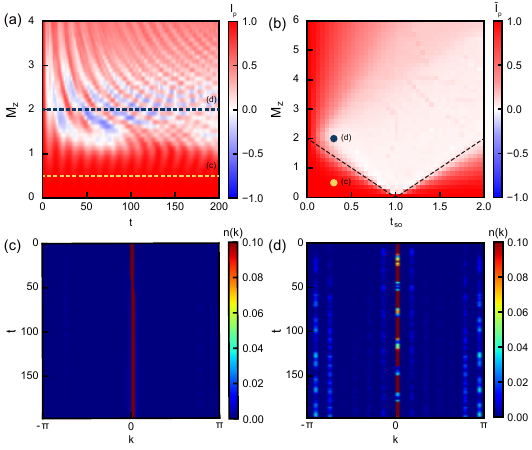}
	\caption{\textbf{Evolution of spin polarization for different phases.} (a) Spin dynamics as a function of $M_z$, with $t_{so}=0.3$, and an initial distribution in the spin up state with uniform distribution. (b) Time-averaged spin polarization as a function of $M_z$ and $t_{\text{so}}$. (c, d) Corresponding reconstructed momentum distribution during the time evolution at extended phase ($M_z=0.5$) and critical phase ($M_z=2.0$). All simulations are conducted with $L=400$ and $\beta=\cos{52^\circ}$ under open boundary conditions.}
	\label{Fig:CP_SpinDynamics}
\end{figure}

We begin by examining the initial state in which all lattice sites are occupied by spin-up atoms ($\ket{\psi}=\ket{\uparrow\uparrow\uparrow\cdots\uparrow}$). In the single-particle picture, this state corresponds to a momentum distribution localized at $k=0$. Figure~\ref{Fig:CP_SpinDynamics}(a), presents the time evolution of the spin polarization $I_p = (n_\uparrow-n_\downarrow)/(n_\uparrow+n_\downarrow)$ under this configuration. In the extended phase, spin polarization exhibits minor oscillations with the value close to 1, while pronounced oscillations are observed in both the critical and localized phases. Utilizing time-averaged spin polarization can aid in characterizing the transition between the critical phase and extend phase (Fig.\ref{Fig:CP_SpinDynamics} (b)). Specifically, it approaches zero in the critical phase but remains close to one in the extended phase. To elucidate the origin of the distinct spin dynamics across the extended–critical transition, we simulate the reconstructed momentum distribution during the evolution (Fig.\ref{Fig:CP_SpinDynamics}(c)–(d)). In the extended phase, the momentum distribution remains localized near $k=0$ throughout the evolution. By contrast, in the critical phase, the distribution broadens to momentum far from $k=0$. In the absence of a quasi-periodic lattice, the system is described by the Bloch Hamiltonian~\cite{zhang2018spin, song2018observation, zhao2023two, liu2013manipulating}:

\begin{equation}
	\mathcal{H}(k)=(m_z-2t_0\cos{k})\sigma_z+2t_{\text{so}}\sin{k}~\sigma_y
\end{equation}

\noindent At $k=0$, the off-diagonal term vanishes, leading to the suppression of spin dynamics at this momentum.

Notably, in the present system, the transition between the critical and localized phases coincides with the topological phase boundary~\cite{wang2020realization}. Away from the topological regime, the system exhibits detuned spin oscillations. Although distinctions in spin dynamics do exist between the critical and localized phases, these differences are less pronounced than the clear demarcation observed between the extended and critical phases. Consequently, spin dynamics may not serve as a reliable indicator for identifying the critical‑to‑localized phase transition.

\begin{figure}[htbp]
	\centering
	\includegraphics[width=1.0\columnwidth]{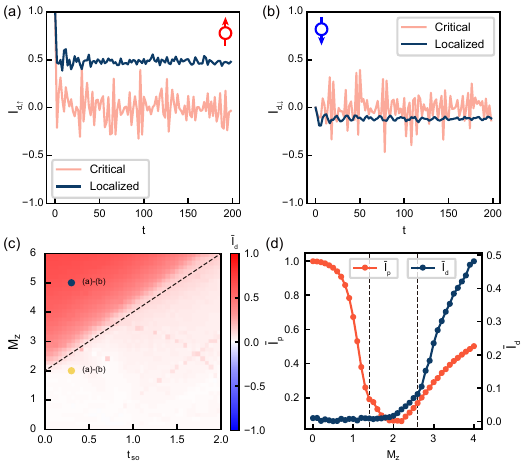}
	\caption{\textbf{Evolution of density imbalance for different phases.} (a, b) Time evolution of the density imbalance for spin-up (a) and spin-down (b) atoms in the critical phase ($M_z=2.0, t_{\text{so}}=0.3$) and the localized phase ($M_z=5.0, t_{\text{so}}=0.3$). (c) Time-averaged density imbalance as a function of $M_z$ and $t_{\text{so}}$. The markers indicate the parameter sets corresponding to the dynamics shown in (a) and (b). (d) Time-averaged spin polarization $\bar{I}_p$ and density imbalance $\bar{I}_d$ as a function of $M_z$ averaged over the time interval $t=0-200$. The expected phase transition position is marked by a vertical dashed line. All simulations are conducted with $L=400$ and $\beta=\cos{52^\circ}$ under open boundary conditions.}
	\label{Fig:CP_DensityImbalance}
\end{figure}

To further probe the critical‑to‑localized phase transition, we examine an alternative initial state in which spin-up atoms half-fill the lattice, described by $\ket{\psi}=\ket{0\uparrow0\uparrow0\uparrow\cdots}$. This state can be experimentally prepared and detected using an optical superlattice ~\cite{schreiber2015observation, luschen2018single}. Figure \ref{Fig:CP_DensityImbalance} (a) and (b) display the time evolution of the density imbalance between even and odd sites, defined as $I_{d, \sigma}=(N_{e, \sigma}-N_{o,\sigma})/(N_{e}+N_{o})$, where $\sigma=\uparrow, \downarrow$ and $N_e (N_o)$ denotes the atom number on even (odd) sites, in both the critical and localized regimes. After long-time evolution, the density imbalance for spin-up atoms remains nonzero in the localized regime, whereas it decays to zero with oscillation in the critical regime. Interestingly, spin-down atoms tend to occupy lattice sites of the opposite parity, as indicated by the opposite sign of their density imbalance. This behavior may originate from the fact that the spin-flip term in our Hamiltonian includes only nearest-neighbor couplings without on-site spin flipping. We can further define the overall density imbalance as $I_d = I_{d,\uparrow}-I_{d, \downarrow}$ and show its time average $\bar{I}_d$ over the interval $t=0-200$ in Fig.~\ref{Fig:CP_DensityImbalance} (c).  A clear boundary emerges between the localized and critical phases. Combined with the time-averaged spin polarization $\bar{I}_p$ introduced earlier, both phase transitions can be clearly identified (Fig.~\ref{Fig:CP_DensityImbalance}), even in systems lacking the ability to prepare or detect Gaussian wavepackets localized to only a few lattice sites.

\section{Coexistence of different states}

\begin{figure}[htbp]
	\centering
	\includegraphics[width=1.0\columnwidth]{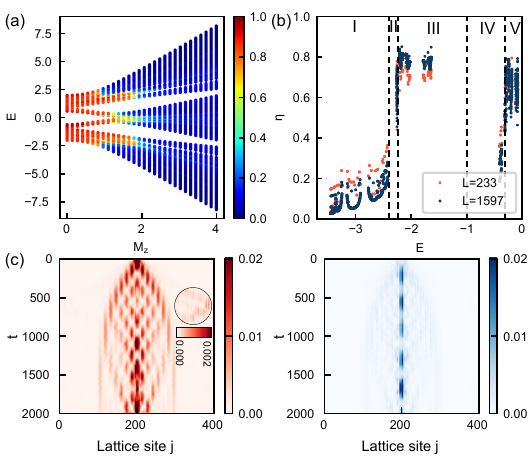}
	\caption{\textbf{Phase with coexisting localized, extended, and critical zones.} (a) Fractal dimension $\eta$ and energy of individual states for distinct values of $M_z$ with $t_{so}=0.3$ and $M_{z,\uparrow}/M_{z,\downarrow}=-2$ with $t_{so}=0.3$. Simulations are conducted with $L=1597$ and $\beta=987/1597$ under periodic boundary condition. (b) Fractal dimension of different eigenstates as a function of eigenenergy for different system size $L=233$ and $L=1597$ with $M_z=1.5$ and $t_{so}=0.3$. For convenience, only the region with $E < 0$ is presented , which is symmetric to that with $E > 0$. (c) Time evolution of the wave packet for the system located in the phase with coexisting localized, extended, and critical zones ($M_z=1.5$, $t_{so}=0.3$ and $M_{z,\uparrow}/M_{z,\downarrow}=-2$). Simulations are conducted with $L=400$ and $\beta=\cos{52^\circ}$ under open boundary condition.}
	\label{Fig:CP_CoexistingPhase}
\end{figure}

Returning to the spin-independent incommensurate lattice case, although the critical phase is absent in such a system, under specific conditions, extended and localized states can coexist, separated by a mobility edge. For a more general situation, where the incommensurate Zeeman potential exhibits a blend of spin-dependence and independence, a phase emerges encompassing coexisting localized, extended, and critical regions~\cite{wang2022quantum}. Illustrated in Fig.\ref{Fig:CP_CoexistingPhase}(a) is the fractal dimension corresponding to various eigenenergies and $M_z$, revealing regions where three or two distinct types of eigenstates coexist, separated by mobility edges. To validate the coexistence of different phases within this domain, we consider the case of $M_z=1.5$ as an example (refer to Fig.\ref{Fig:CP_CoexistingPhase}(b)). Notably, as the system size increases, $\eta$ converges towards 0 for all states in zones I and IV, indicative of localization states, while tending towards 1 for all states in zone III, signifying extended states. Conversely, within zones II and V, the fractal dimension markedly deviates from 0 and 1 and remains relatively constant with system size, suggesting a critical state in these regions. 

The time evolution of the wave packet under such circumstances, as depicted in Fig.\ref{Fig:CP_CoexistingPhase}(c), showcases characteristic signatures of these three distinct phases concurrently. On the whole, the time evolution of the wave packet exhibits quasi-localization, highlighting the presence of critical states. However, the higher concentration of atom clouds in the vicinity of central sites, particularly for spin-down atoms, indicates the existence of localized states. Furthermore, the inset of Fig.\ref{Fig:CP_CoexistingPhase}(c) illustrates instances of ballistic expansion behavior of localized states upon closer inspection of the color bar. Through simulations of time evolution, the coexistence of extended, critical, and localized states under such conditions is further substantiated.

\section{Localization behavior in non-Hermitian regime}

Recently, various non-Hermitian extensions of systems exhibiting Anderson localization have been explored, such as the non-reciprocal Aubry-André model~\cite{Jiang2019r9s} and the PT-symmetric Aubry-André model~\cite{yuce2014pt, liang2014pt, hang2015localization, harter2016pt, zeng2017anderson, longhi2019topological}. Numerical studies have delved into the impact of nonreciprocal tunneling, physical gain and loss on parity-time symmetry breaking, the Hofstadter butterfly spectrum~\cite{longhi2014PT, yuce2014pt, liang2014pt, harter2016pt}, topological phase transitions~\cite{longhi2019topological}, and the localization characteristics of eigenstates~\cite{hatano1996localization, shnerb1998winding, hang2015localization, zeng2017anderson, Jiang2019r9s, zhou2023non}. Given these insights, it becomes pertinent to investigate the effects of non-Hermiticity on our system.

\begin{figure}[htbp]
	\centering
	\includegraphics[width=1.0\columnwidth]{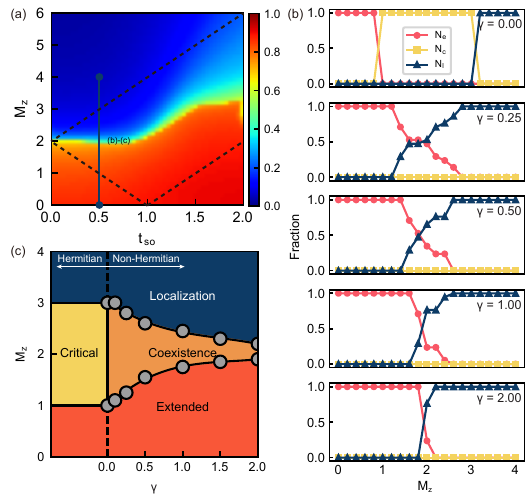}
	\caption{\textbf{Suppression of critical phase in non-Hermitian Raman lattice.} (a) Mean fractal dimension $\bar{\eta}$ of the system under a non-Hermtian term ($\gamma=2$) when $M_{z,\uparrow}/M_{z,\downarrow}=-1$. $M_z$ and $t_{so}$ are presented in units of $t_0$. The dashed lines show the boundary between different phases in Hermitian regime. (b) Fraction of eigenstates belonging to the extended, critical, and localized phases as a function of $M_z$ for different values of $\gamma$ at $t_{\text{so}}=0.5$. In the non-Hermitian regime ($\gamma\ne0$), the critical phase is significantly suppressed. (c) As the system enters the non-Hermitian regime, the critical phase evolves into a mixed phase with coexisting localized and extended states. The purely localized and extended phases are also enlarged with increasing non-Hermitian strength.}
	\label{Fig:CP_Non-Hermitian}
\end{figure}

Expanding our setup into the non-Hermitian regime involves introducing spin-dependent dissipation to one of the spin states ($\gamma_\sigma\ne0$) through a nearly resonant loss beam~\cite{zhou2023non}. In such system, one generally focus on the post-selected space, which corresponds to the remaining atoms. Under this condition, both the eigenspectrum and the time evolution can be well descibed by the effective non-Hermtian Hamiltonian~\cite{li2022engineering, Ashida2021}. Without loss of the generality, in the therotical simulation, we consider a system with balanced gain and loss term ($\gamma_\uparrow=-\gamma_\downarrow=\gamma$).  We focus on the scenario with $M_{z,\uparrow}/M_{z,\downarrow}=-1$, which exhibits a critical phase in the Hermitian context. The mean fractal dimension for the system with non-Hermitian term ($\gamma=2$) is illustrated in Fig. \ref{Fig:CP_Non-Hermitian}(a). The region with $0<\bar{\eta}<1$ vanishes, potentially indicating that the critical phases in the Hermitian regime is suppressed. For system
with larger $M_z$, the mean fractal dimension decreases to a near zero value, indicative of a transformation of eigenstates into localized states, whereas for lower $M_z$ values, the mean fractal dimension rises to near one value, signifying a transformation towards extended states. To avoid the finite size effect, we further present the fraction of extended, critical and localized phases for varying dissipation values in Fig.~\ref{Fig:CP_Non-Hermitian}(b) with $t_\text{so}=0.5$. In the non-Hermitian regime ($\gamma\ne0$), the critical phase is significantly suppressed, and the critical phase changed into a mixed phase with coexisting localized and extended
states. Furthermore, we can also find as the non-Hermitian term increases, the pure localized and extended phases are also enlarged (Fig.~\ref{Fig:CP_Non-Hermitian} (c)).

The mechanism for the emergence of critical states in the Hermitian regime of this spinful model also provides insight for their suppression under non-Hermitian perturbations. As revealed in Refs.\cite{zhou2023exact, zhou2025fundamental}, critical states in such systems arise when the on-site potential exhibits generalized incommensurate zeros in its $\sigma_z$ component. These zeros can be mapped to incommensurately distributed zeros in the hopping coefficients of an effective one-dimensional spinless model. The presence of such zeros effectively fragments the system into decoupled segments, forcing the delocalized wave functions to reorganize and develop the characteristic self-similar structure of critical states. However, the introduction of non-Hermitian terms disrupts the condition for generalized incommensurate zeros in the on-site potential. Consequently, the formation of critical states is invariably suppressed.

\vspace{10pt}
\section{Conclusion and outlook}

This study proposes a versatile Raman lattice configuration for theoretically investigating localization behavior for alkaline-earth-like atoms, using $^{173}$Yb as an example. Unlike previous approaches that primarily employed alkali-metal atoms~\cite{wang2020realization, wang2022quantum, zhou2025fundamental}, $^{173}$Yb represents a practically demonstrated fermionic system in optical Raman lattices~\cite{song2018observation, Song2019, zhao2023two}. The fermionic nature of $^{173}$Yb, along with the potential SU(N) symmetry of alkaline-earth atoms, can give rise to phenomena distinct from bosonic systems, particularly in the many-body regime~\cite{molignini2025stability2, wang2021many, wang2020realization}.

In this theoretical framework, we explore three distinct quasi-periodic Raman lattice configurations: an entirely spin-dependent incommensurate lattice in a Hermitian regime, a partially spin-dependent incommensurate lattice under Hermitian conditions, and a fully spin-dependent incommensurate lattice in a non-Hermitian regime. In the case of the entirely spin-dependent incommensurate lattice in the Hermitian regime, our model predicts the emergence of a critical phase, which can be theoretically detected through signatures in expansion dynamics and spin polarization evolution. Conversely, when the incommensurate lattice transitions into a combination of spin-dependent and spin-independent characteristics, theoretical calculations reveal a new phase with coexistence of extended, critical, and localized phases separated by mobility edges. Furthermore, we demonstrate that introducing non-Hermitian dissipation disrupts the generalized incommensurate zeros and further suppresses the critical phase. This work bridges a gap by providing a practical experimental scheme for implementing and studying localization with fermionic alkaline-earth atoms in optical Raman lattices, directly addressing experimental capabilities. The proposed scheme also facilitates exploration of the interplay between localization behavior and non-Hermiticity, thereby laying a foundation for future experimental studies with ultracold atoms~\cite{zhou2023non,Jiang2019r9s}.

\begin{acknowledgments}
We thank Xin-Chi Zhou and Xiong-Jun Liu for fruitful discussions. GBJ acknowledges support from the RGC through 16302123, 16305024 and RFS2122-6S04. 
\end{acknowledgments}


\appendix

\section{Finite size analysis}

    \begin{figure*}[htbp]
	\centering
	\includegraphics[width=0.9\linewidth]{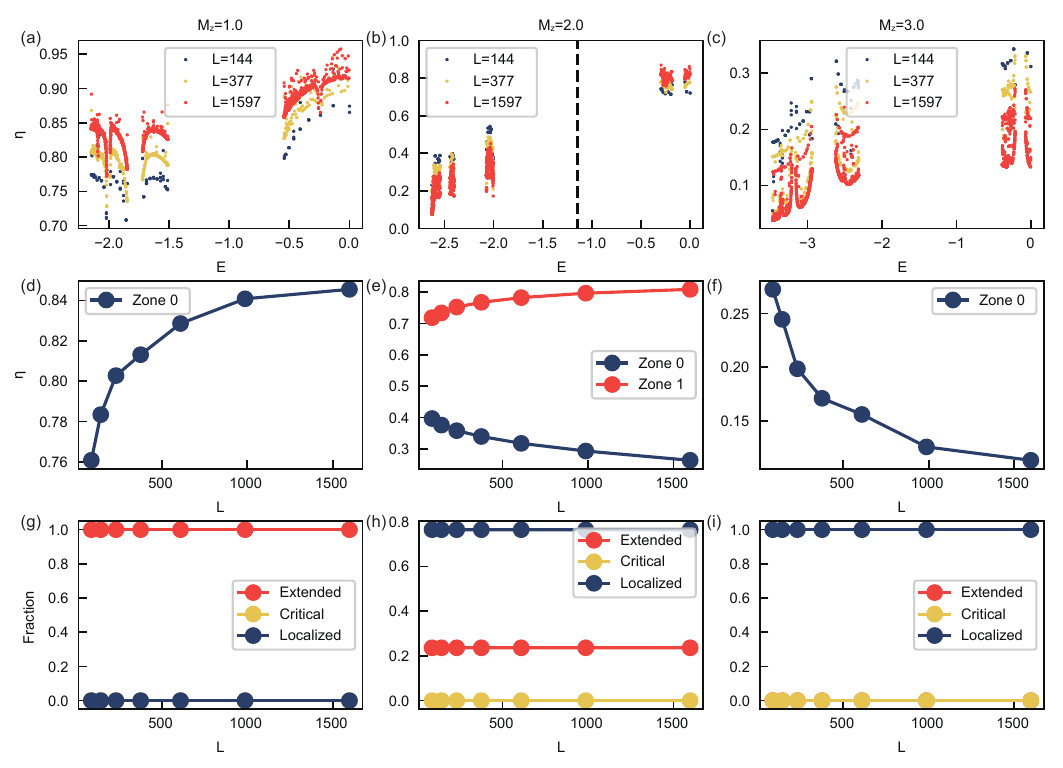}
	\caption{\textbf{Finite size effects on $N_e, N_c, N_l$ and $\eta$.} (a)-(c) Fractal dimension $\eta$ of individual eigenstates computed at three system sizes $L$, corresponding to $M_z=1.0, 2.0, 3.0$ respectively. (d)-(f) Scaling of $\eta$ with system size $L$ across different zones. (g)-(i) Fractions of extended, critical, and localized states ($N_e/N, N_c/N, N_l/N$) as functions of system size $L$. All results are simulated with $\gamma=1.0$ and $t_{\text{so}}=0.5$ in periodic boundary conditions.}
	\label{Fig:SM_FiniteSizeAnalysis}
    \end{figure*} 

    In this section, we investigate the finite size effects on the fractal dimension and the fractions of extended, critical, and localized states ($N_e/N, N_c/N, N_l/N$) in the non-Hermitian regime. Figure~\ref{Fig:SM_FiniteSizeAnalysis}(a)-(c) shows the fractal dimension $\eta$ for $t_\text{so}=0.5$ and $\gamma=1.0$ at different system sizes $L$. For lower $M_z$ (e.g., $M_z=1.0$), the fractal dimensions of all eigenstates increase with system size and approach 1, indicating extended behavior. Conversely, for higher $M_z$ (e.g., $M_z=3.0$), the fractal dimensions decrease with increasing $L$ and tend to zero, consistent with behaviors of localized states. In the intermediate regime ($M_z=2.0$), eigenstates with larger absolute eigenenergies ($|E|>1.15$) exhibit decreasing fractal dimensions as $L$ increases, while those with lower energies show the opposite trend, suggesting a phase with coexisting extended and localized states.

    Based on these observations, we classify eigenstates into distinct zones sharing similar localization properties. The average fractal dimension within each zone is presented in Figure~\ref{Fig:SM_FiniteSizeAnalysis}(d)-(f). For $M_z=1.0$ and $3.0$, all eigenstates fall into a single zone, corresponding to a pure phase. At $M_z=2.0$, however, eigenstates separate into two zones with mean fractal dimensions evolving oppositely with system size—one increasing and the other decreasing. This allows us to assign consistent localization properties to all states within each zone and determine the fractions of extended, critical, and localized states. As shown in Figure~\ref{Fig:SM_FiniteSizeAnalysis}(g)-(i), these fractions remain invariant with increasing system size.

    \section{Absence of non-Hermitian skin effect}

\begin{figure*}[htbp]
	\centering
	\includegraphics[width=0.9\linewidth]{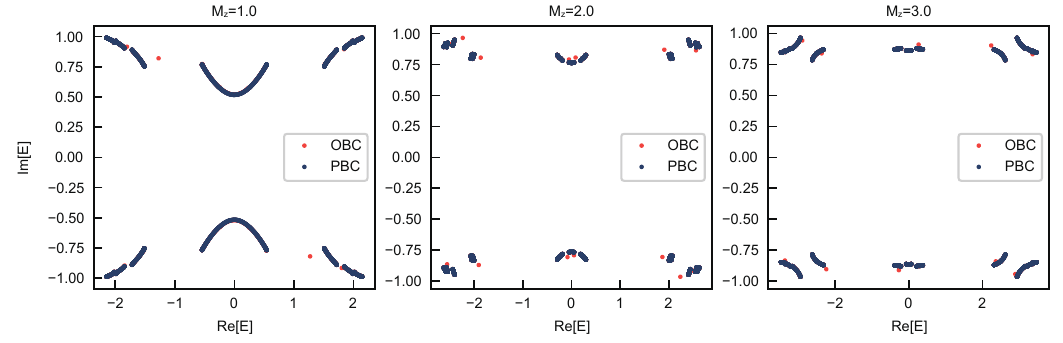}
	\caption{\textbf{Eigenenergy spectrum on complex plane under PBC and OBC with $M_z=1.0, 2.0, 3.0$ respectively.} All results are simulated with $\gamma=1.0$ and $t_{\text{so}}=0.5$.}
	\label{Fig:SM_SkinEffect}
    \end{figure*}

    In this section, we validate the absence of the non-Hermitian skin effect in our system, confirming that the observed suppression of the critical phase does not originate from the skin effect. According to the established correspondence between winding numbers and skin modes in non-Hermitian systems~\cite{zhang2020correspondence, zhao2023two, okuma2020topological, zhou2022engineering}, a system exhibits the skin effect under open boundary conditions (OBC) if and only if the eigenvalue winding on the complex plane displays nontrivial closed-loop topology under periodic boundary conditions (PBC). The presence of the skin effect also leads to distinct eigenspectra under PBC and OBC. As shown in Figure~\ref{Fig:SM_SkinEffect}, the eigenenergy spectra on the complex plane are presented for both boundary conditions. The PBC spectrum exhibits no loop structures, indicating the absence of the non-Hermitian skin effect in the present system.

    \section{Effect of the width of the initial wave packet on the expansion dynamics}

    \begin{figure*}[htbp]
	\centering
	\includegraphics[width=0.875\linewidth]{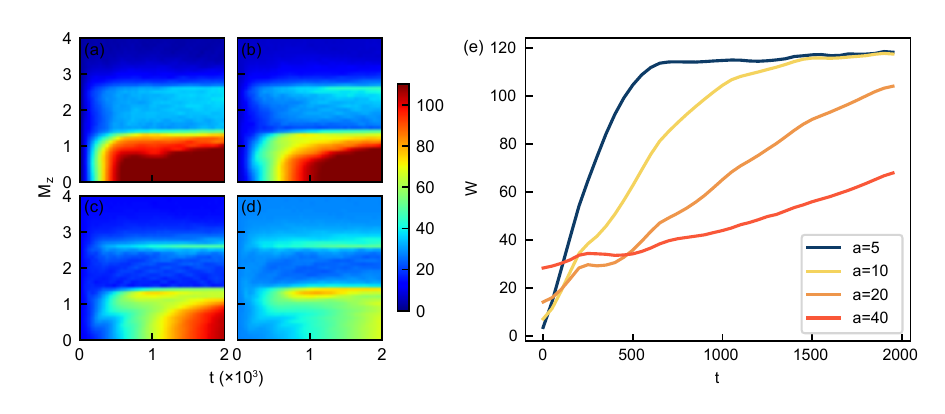}
	\caption{\textbf{Effect of the width of initial wave packet.} (a-d) Expansion dynamics of the wave packet, characterized by $W$ as a function of $M_z$, with $t_{so}=0.3$ and an initial width of $a=5, 10, 20, 40$, respectively. (e) $W$ as a function of time $t$ for different widths with a fixed $M_z=0.5$. Other parameters are the same as the main Fig. 3.}
	\label{Fig:SM_WavePacket}
    \end{figure*} 

    \begin{figure*}[htbp]
	\centering
	\includegraphics[width=0.95\linewidth]{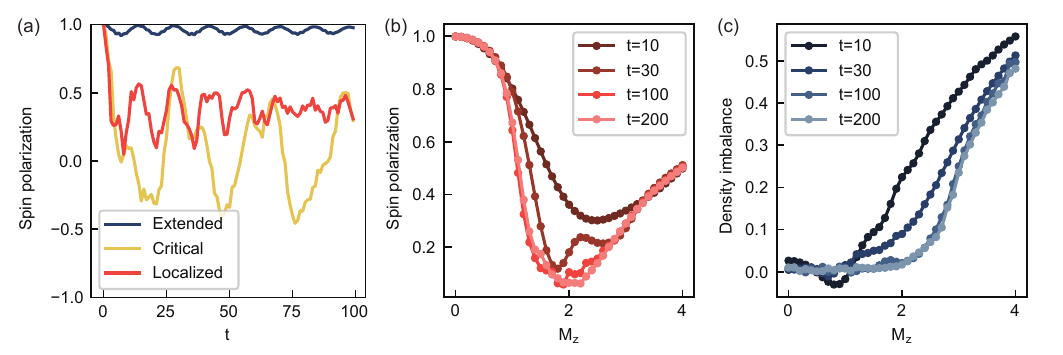}
	\caption{\textbf{Effect of the total evolution time.} (a) Time evolution of the spin polarization for the system located in the extended phase ($M_z$ = 0.5), critical phase ($M_z$ = 2.0), and localized phase ($M_z$ = 3.5) with the initial state comprising spin-up atoms in a uniform distribution.(b) Time-averaged spin polarization $\bar{I}_p$ as a function of $M_z$ averaged over different time interval $t_f=10, 30, 100, 200$.(c) Time-averaged density imbalance $\bar{I}_d$ as a function of $M_z$ averaged over different time interval $t_f=10, 30, 100, 200$. Other parameters are identical to those in Fig.5(d). }
	\label{Fig:SM_TimeEvolution}
    \end{figure*} 

    In this section, we investigate the effect of the initial wave packet width $a$ on the expansion dynamics. We find that while the initial wave packet width $a$ does not alter the qualitative distinction in dynamical behaviors among different phases, it does quantitatively influence the clarity of the contrast and the expansion speed, as shown in Figure~\ref{Fig:SM_WavePacket}.\\

    \section{Effect of the evolution time on time average $\bar{I}_d$ and $\bar{I}_p$}

    In this section, we examine the influence of the total evolution time on the dynamics of the density imbalance and spin polarization. We observe that while a very short evolution time can quantitatively affect the time-averaged spin polarization and density imbalance, the qualitative trends across different phases remain unchanged, as demonstrated in Figure~\ref{Fig:SM_TimeEvolution}. Moreover, the time-averaged spin polarization is robust against the total evolution time once it spans several oscillation periods. Hence, despite experimental constraints that may limit the coherent evolution time, spin dynamics can still serve as a reliable detection scheme for distinguishing between phases—provided the total evolution time is sufficiently long to encompass several periods of spin oscillation.\\

\bibliography{ref.bib}

@article{liu2025, 
year = {2025}, 
title = {{Third-order exceptional point in non-Hermitian spin-orbit-coupled cold atoms}}, 
author = {Liu, Yu-Jun and Pak, Ka Kwan and Ren, Peng and Guo, Mengbo and Zhao, Entong and He, Chengdong and Jo, Gyu-Boong}, 
journal = {Physical Review A}, 
pages = {023305}, 
number = {2}, 
volume = {112}
}

@article{anderson1958absence,
	title={Absence of diffusion in certain random lattices},
	author={Anderson, Philip W},
	journal={Physical review},
	volume={109},
	number={5},
	pages={1492},
	year={1958},
	publisher={APS}
}

@article{thouless1974electrons,
	title={Electrons in disordered systems and the theory of localization},
	author={Thouless, David J},
	journal={Physics Reports},
	volume={13},
	number={3},
	pages={93--142},
	year={1974},
	publisher={Elsevier}
}

@article{abrahams1979scaling,
	title={Scaling theory of localization: Absence of quantum diffusion in two dimensions},
	author={Abrahams, Elihu and Anderson, Philip W and Licciardello, Donald C and Ramakrishnan, Tiruppattur V},
	journal={Physical Review Letters},
	volume={42},
	number={10},
	pages={673},
	year={1979},
	publisher={APS}
}

@article{grempel1982localization,
	title={Localization in an incommensurate potential: An exactly solvable model},
	author={Grempel, DR and Fishman, Shmuel and Prange, RE},
	journal={Physical Review Letters},
	volume={49},
	number={11},
	pages={833},
	year={1982},
	publisher={APS}
}

@article{halsey1986fractal,
	title={Fractal measures and their singularities: The characterization of strange sets},
	author={Halsey, Thomas C and Jensen, Mogens H and Kadanoff, Leo P and Procaccia, Itamar and Shraiman, Boris I},
	journal={Physical review A},
	volume={33},
	number={2},
	pages={1141},
	year={1986},
	publisher={APS}
}

@article{machida1986quantum,
	title={Quantum energy spectra and one-dimensional quasiperiodic systems},
	author={Machida, Kazushige and Fujita, Mitsutaka},
	journal={Physical Review B},
	volume={34},
	number={10},
	pages={7367},
	year={1986},
	publisher={APS}
}

@article{hisashi1988dynamics,
	title={Dynamics of an Electron in Quasiperiodic Systems. II. Harper’s Model},
	author={Hisashi, Hiramoto and Shuji, Abe},
	journal={Journal of the Physical Society of Japan},
	volume={57},
	number={4},
	pages={1365--1371},
	year={1988},
	publisher={一般社団法人 日本物理学会}
}

@article{sarma1988mobility,
	title={Mobility edge in a model one-dimensional potential},
	author={Sarma, S Das and He, Song and Xie, XC},
	journal={Physical review letters},
	volume={61},
	number={18},
	pages={2144},
	year={1988},
	publisher={APS}
}

@article{geisel1991new,
	title={New class of level statistics in quantum systems with unbounded diffusion},
	author={Geisel, T and Ketzmerick, R and Petschel, G},
	journal={Physical review letters},
	volume={66},
	number={13},
	pages={1651},
	year={1991},
	publisher={APS}
}

@article{ketzmerick1997determines,
	title={What determines the spreading of a wave packet?},
	author={Ketzmerick, R and Kruse, K and Kraut, S and Geisel, T},
	journal={Physical review letters},
	volume={79},
	number={11},
	pages={1959},
	year={1997},
	publisher={APS}
}

@article{aulbach2004phase,
	title={Phase-space visualization of a metal--insulator transition},
	author={Aulbach, Christian and Wobst, Andr{\'e} and Ingold, Gert-Ludwig and H{\"a}nggi, Peter and Varga, Imre},
	journal={New Journal of Physics},
	volume={6},
	number={1},
	pages={70},
	year={2004},
	publisher={IOP Publishing}
}

@article{mirlin2006exact,
	title={Exact relations between multifractal exponents at the Anderson transition},
	author={Mirlin, Alexander D and Fyodorov, Yu V and Mildenberger, A and Evers, Ferdinand},
	journal={Physical review letters},
	volume={97},
	number={4},
	pages={046803},
	year={2006},
	publisher={APS}
}

@article{evers2008anderson,
	title={Anderson transitions},
	author={Evers, Ferdinand and Mirlin, Alexander D},
	journal={Reviews of Modern Physics},
	volume={80},
	number={4},
	pages={1355--1417},
	year={2008},
	publisher={APS}
}

@article{billy2008direct,
	title={Direct observation of Anderson localization of matter waves in a controlled disorder},
	author={Billy, Juliette and Josse, Vincent and Zuo, Zhanchun and Bernard, Alain and Hambrecht, Ben and Lugan, Pierre and Cl{\'e}ment, David and Sanchez-Palencia, Laurent and Bouyer, Philippe and Aspect, Alain},
	journal={Nature},
	volume={453},
	number={7197},
	pages={891--894},
	year={2008},
	publisher={Nature Publishing Group}
}

@article{roati2008anderson,
	title={Anderson localization of a non-interacting Bose--Einstein condensate},
	author={Roati, Giacomo and D’Errico, Chiara and Fallani, Leonardo and Fattori, Marco and Fort, Chiara and Zaccanti, Matteo and Modugno, Giovanni and Modugno, Michele and Inguscio, Massimo},
	journal={Nature},
	volume={453},
	number={7197},
	pages={895--898},
	year={2008},
	publisher={Nature Publishing Group UK London}
}

@article{bakr2009quantum,
	title={A quantum gas microscope for detecting single atoms in a Hubbard-regime optical lattice},
	author={Bakr, Waseem S and Gillen, Jonathon I and Peng, Amy and F{\"o}lling, Simon and Greiner, Markus},
	journal={Nature},
	volume={462},
	number={7269},
	pages={74--77},
	year={2009},
	publisher={Nature Publishing Group UK London}
}

@article{larcher2009effects,
	title={Effects of interaction on the diffusion of atomic matter waves in one-dimensional quasiperiodic potentials},
	author={Larcher, Marco and Dalfovo, Franco and Modugno, Michele},
	journal={Physical Review A},
	volume={80},
	number={5},
	pages={053606},
	year={2009},
	publisher={APS}
}

@article{biddle2010predicted,
	title={Predicted Mobility Edges in One-Dimensional Incommensurate Optical Lattices:<? format?> An Exactly Solvable Model of Anderson Localization},
	author={Biddle, J and Das Sarma, S},
	journal={Physical review letters},
	volume={104},
	number={7},
	pages={070601},
	year={2010},
	publisher={APS}
}

@article{sherson2010single,
	title={Single-atom-resolved fluorescence imaging of an atomic Mott insulator},
	author={Sherson, Jacob F and Weitenberg, Christof and Endres, Manuel and Cheneau, Marc and Bloch, Immanuel and Kuhr, Stefan},
	journal={Nature},
	volume={467},
	number={7311},
	pages={68--72},
	year={2010},
	publisher={Nature Publishing Group UK London}
}

@article{liu2013manipulating,
	title={Manipulating topological edge spins in a one-dimensional optical lattice},
	author={Liu, Xiong-Jun and Liu, Zheng-Xin and Cheng, Meng},
	journal={Physical review letters},
	volume={110},
	number={7},
	pages={076401},
	year={2013},
	publisher={APS}
}

@article{dubertrand2014two,
	title={Two scenarios for quantum multifractality breakdown},
	author={Dubertrand, R{\'e}my and Garc{\'\i}a-Mata, Ignacio and Georgeot, Bertrand and Giraud, Olivier and Lemari{\'e}, Gabriel and Martin, John},
	journal={Physical Review Letters},
	volume={112},
	number={23},
	pages={234101},
	year={2014},
	publisher={APS}
}

@article{murthy2014matter,
	title={Matter-wave Fourier optics with a strongly interacting two-dimensional Fermi gas},
	author={Murthy, PA and Kedar, D and Lompe, T and Neidig, M and Ries, MG and Wenz, AN and Z{\"u}rn, G and Jochim, S},
	journal={Physical Review A},
	volume={90},
	number={4},
	pages={043611},
	year={2014},
	publisher={APS}
}

@article{edge2015imaging,
	title={Imaging and addressing of individual fermionic atoms in an optical lattice},
	author={Edge, Graham JA and Anderson, Rhys and Jervis, Dylan and McKay, David C and Day, Ryan and Trotzky, Stefan and Thywissen, Joseph H},
	journal={Physical Review A},
	volume={92},
	number={6},
	pages={063406},
	year={2015},
	publisher={APS}
}

@article{miranda2015site,
	title={Site-resolved imaging of ytterbium atoms in a two-dimensional optical lattice},
	author={Miranda, Martin and Inoue, Ryotaro and Okuyama, Yuki and Nakamoto, Akimasa and Kozuma, Mikio},
	journal={Physical Review A},
	volume={91},
	number={6},
	pages={063414},
	year={2015},
	publisher={APS}
}

@article{haller2015single,
	title={Single-atom imaging of fermions in a quantum-gas microscope},
	author={Haller, Elmar and Hudson, James and Kelly, Andrew and Cotta, Dylan A and Peaudecerf, Bruno and Bruce, Graham D and Kuhr, Stefan},
	journal={Nature Physics},
	volume={11},
	number={9},
	pages={738--742},
	year={2015},
	publisher={Nature Publishing Group UK London}
}

@article{omran2015microscopic,
	title={Microscopic observation of Pauli blocking in degenerate fermionic lattice gases},
	author={Omran, Ahmed and Boll, Martin and Hilker, Timon A and Kleinlein, Katharina and Salomon, Guillaume and Bloch, Immanuel and Gross, Christian},
	journal={Physical review letters},
	volume={115},
	number={26},
	pages={263001},
	year={2015},
	publisher={APS}
}

@article{zhen2015spawning,
	title={Spawning rings of exceptional points out of Dirac cones},
	author={Zhen, Bo and Hsu, Chia Wei and Igarashi, Yuichi and Lu, Ling and Kaminer, Ido and Pick, Adi and Chua, Song-Liang and Joannopoulos, John D and Solja{\v{c}}i{\'c}, Marin},
	journal={Nature},
	volume={525},
	number={7569},
	pages={354--358},
	year={2015},
	publisher={Nature Publishing Group UK London}
}

@article{bertrand2016anomalous,
	title={Anomalous Thouless energy and critical statistics on the metallic side of the many-body localization transition},
	author={Bertrand, Corentin L and Garc{\'\i}a-Garc{\'\i}a, Antonio M},
	journal={Physical Review B},
	volume={94},
	number={14},
	pages={144201},
	year={2016},
	publisher={APS}
}

@article{cheuk2016observation,
	title={Observation of 2D fermionic Mott insulators of K 40 with single-site resolution},
	author={Cheuk, Lawrence W and Nichols, Matthew A and Lawrence, Katherine R and Okan, Melih and Zhang, Hao and Zwierlein, Martin W},
	journal={Physical review letters},
	volume={116},
	number={23},
	pages={235301},
	year={2016},
	publisher={APS}
}

@article{parsons2016site,
	title={Site-resolved measurement of the spin-correlation function in the Fermi-Hubbard model},
	author={Parsons, Maxwell F and Mazurenko, Anton and Chiu, Christie S and Ji, Geoffrey and Greif, Daniel and Greiner, Markus},
	journal={Science},
	volume={353},
	number={6305},
	pages={1253--1256},
	year={2016},
	publisher={American Association for the Advancement of Science}
}

@article{yamamoto2016ytterbium,
	title={An ytterbium quantum gas microscope with narrow-line laser cooling},
	author={Yamamoto, Ryuta and Kobayashi, Jun and Kuno, Takuma and Kato, Kohei and Takahashi, Yoshiro},
	journal={New Journal of Physics},
	volume={18},
	number={2},
	pages={023016},
	year={2016},
	publisher={IOP Publishing}
}

@article{zeng2017anderson,
	title={Anderson localization in the non-Hermitian Aubry-Andr{\'e}-Harper model with physical gain and loss},
	author={Zeng, Qi-Bo and Chen, Shu and L{\"u}, Rong},
	journal={Physical Review A},
	volume={95},
	number={6},
	pages={062118},
	year={2017},
	publisher={APS}
}

@article{song2018observation,
	title={Observation of symmetry-protected topological band with ultracold fermions},
	author={Song, Bo and Zhang, Long and He, Chengdong and Poon, Ting Fung Jeffrey and Hajiyev, Elnur and Zhang, Shanchao and Liu, Xiong-Jun and Jo, Gyu-Boong},
	journal={Science advances},
	volume={4},
	number={2},
	pages={eaao4748},
	year={2018},
	publisher={American Association for the Advancement of Science}
}

@article{mitra2018quantum,
	title={Quantum gas microscopy of an attractive Fermi--Hubbard system},
	author={Mitra, Debayan and Brown, Peter T and Guardado-Sanchez, Elmer and Kondov, Stanimir S and Devakul, Trithep and Huse, David A and Schau{\ss}, Peter and Bakr, Waseem S},
	journal={Nature Physics},
	volume={14},
	number={2},
	pages={173--177},
	year={2018},
	publisher={Nature Publishing Group UK London}
}

@article{luschen2018single,
	title={Single-particle mobility edge in a one-dimensional quasiperiodic optical lattice},
	author={L{\"u}schen, Henrik P and Scherg, Sebastian and Kohlert, Thomas and Schreiber, Michael and Bordia, Pranjal and Li, Xiao and Das Sarma, S and Bloch, Immanuel},
	journal={Physical review letters},
	volume={120},
	number={16},
	pages={160404},
	year={2018},
	publisher={APS}
}

@article{an2018engineering,
	title={Engineering a flux-dependent mobility edge in disordered zigzag chains},
	author={An, Fangzhao Alex and Meier, Eric J and Gadway, Bryce},
	journal={Physical Review X},
	volume={8},
	number={3},
	pages={031045},
	year={2018},
	publisher={APS}
}

@article{zhou2018observation,
	title={Observation of bulk Fermi arc and polarization half charge from paired exceptional points},
	author={Zhou, Hengyun and Peng, Chao and Yoon, Yoseob and Hsu, Chia Wei and Nelson, Keith A and Fu, Liang and Joannopoulos, John D and Solja{\v{c}}i{\'c}, Marin and Zhen, Bo},
	journal={Science},
	volume={359},
	number={6379},
	pages={1009--1012},
	year={2018},
	publisher={American Association for the Advancement of Science}
}

@incollection{zhang2018spin,
	title={Spin-orbit coupling and topological phases for ultracold atoms},
	author={Zhang, Long and Liu, Xiong-Jun},
	booktitle={Synthetic Spin-Orbit Coupling in Cold Atoms},
	pages={1--87},
	year={2018},
	publisher={World Scientific}
}

@article{wang2018dirac,
	title={Dirac-, Rashba-, and Weyl-type spin-orbit couplings: Toward experimental realization in ultracold atoms},
	author={Wang, Bao-Zong and Lu, Yue-Hui and Sun, Wei and Chen, Shuai and Deng, Youjin and Liu, Xiong-Jun},
	journal={Physical Review A},
	volume={97},
	number={1},
	pages={011605},
	year={2018},
	publisher={APS}
}

@article{ezawa2019non,
	title={Non-Hermitian higher-order topological states in nonreciprocal and reciprocal systems with their electric-circuit realization},
	author={Ezawa, Motohiko},
	journal={Physical Review B},
	volume={99},
	number={20},
	pages={201411},
	year={2019},
	publisher={APS}
}

@article{wang2020realization,
	title={Realization and Detection of Nonergodic Critical Phases in an Optical Raman Lattice},
	author={Wang, Yucheng and Zhang, Long and Niu, Sen and Yu, Dapeng and Liu, Xiong-Jun},
	journal={Physical Review Letters},
	volume={125},
	number={7},
	pages={073204},
	year={2020},
	publisher={APS}
}

@article{weidemann2020topological,
	title={Topological funneling of light},
	author={Weidemann, Sebastian and Kremer, Mark and Helbig, Tobias and Hofmann, Tobias and Stegmaier, Alexander and Greiter, Martin and Thomale, Ronny and Szameit, Alexander},
	journal={Science},
	volume={368},
	number={6488},
	pages={311--314},
	year={2020},
	publisher={American Association for the Advancement of Science}
}

@article{helbig2020generalized,
	title={Generalized bulk--boundary correspondence in non-Hermitian topolectrical circuits},
	author={Helbig, Tobias and Hofmann, Tobias and Imhof, S and Abdelghany, M and Kiessling, T and Molenkamp, LW and Lee, CH and Szameit, A and Greiter, M and Thomale, R},
	journal={Nature Physics},
	volume={16},
	number={7},
	pages={747--750},
	year={2020},
	publisher={Nature Publishing Group}
}

@article{xiao2020non,
	title={Non-Hermitian bulk--boundary correspondence in quantum dynamics},
	author={Xiao, Lei and Deng, Tianshu and Wang, Kunkun and Zhu, Gaoyan and Wang, Zhong and Yi, Wei and Xue, Peng},
	journal={Nature Physics},
	volume={16},
	number={7},
	pages={761--766},
	year={2020},
	publisher={Nature Publishing Group}
}

@article{yamamoto2020single,
	title={Single-site-resolved imaging of ultracold atoms in a triangular optical lattice},
	author={Yamamoto, Ryuta and Ozawa, Hideki and Nak, David C and Nakamura, Ippei and Fukuhara, Takeshi},
	journal={New Journal of Physics},
	volume={22},
	number={12},
	pages={123028},
	year={2020},
	publisher={IOP Publishing}
}

@article{zhang2020non,
	title={Non-hermitian topological anderson insulators},
	author={Zhang, Dan-Wei and Tang, Ling-Zhi and Lang, Li-Jun and Yan, Hui and Zhu, Shi-Liang},
	journal={Science China Physics, Mechanics \& Astronomy},
	volume={63},
	number={6},
	pages={267062},
	year={2020},
	publisher={Springer}
}

@article{an2021interactions,
	title={Interactions and mobility edges: Observing the generalized aubry-andr{\'e} model},
	author={An, Fangzhao Alex and Padavi{\'c}, Karmela and Meier, Eric J and Hegde, Suraj and Ganeshan, Sriram and Pixley, JH and Vishveshwara, Smitha and Gadway, Bryce},
	journal={Physical review letters},
	volume={126},
	number={4},
	pages={040603},
	year={2021},
	publisher={APS}
}

@article{asteria2021quantum,
	title={Quantum gas magnifier for sub-lattice-resolved imaging of 3D quantum systems},
	author={Asteria, Luca and Zahn, Henrik P and Kosch, Marcel N and Sengstock, Klaus and Weitenberg, Christof},
	journal={Nature},
	volume={599},
	number={7886},
	pages={571--575},
	year={2021},
	publisher={Nature Publishing Group UK London}
}

@article{xiao2021observation,
	title={Observation of topological phase with critical localization in a quasi-periodic lattice},
	author={Xiao, Teng and Xie, Dizhou and Dong, Zhaoli and Chen, Tao and Yi, Wei and Yan, Bo},
	journal={Science bulletin},
	volume={66},
	number={21},
	pages={2175--2180},
	year={2021},
	publisher={Elsevier}
}

@article{kwon2022site,
	title={Site-resolved imaging of a bosonic Mott insulator of Li 7 atoms},
	author={Kwon, Kiryang and Kim, Kyungtae and Hur, Junhyeok and Huh, SeungJung and Choi, Jae-yoon},
	journal={Physical Review A},
	volume={105},
	number={3},
	pages={033323},
	year={2022},
	publisher={APS}
}

@article{liang2022dynamic,
	title={Dynamic signatures of non-Hermitian skin effect and topology in ultracold atoms},
	author={Liang, Qian and Xie, Dizhou and Dong, Zhaoli and Li, Haowei and Li, Hang and Gadway, Bryce and Yi, Wei and Yan, Bo},
	journal={Physical review letters},
	volume={129},
	number={7},
	pages={070401},
	year={2022},
	publisher={APS}
}

@article{wang2022quantum,
	title={Quantum phase with coexisting localized, extended, and critical zones},
	author={Wang, Yucheng and Zhang, Long and Sun, Wei and Poon, Ting-Fung Jeffrey and Liu, Xiong-Jun},
	journal={Physical Review B},
	volume={106},
	number={14},
	pages={L140203},
	year={2022},
	publisher={APS}
}

@article{ren2022chiral,
	title={Chiral control of quantum states in non-Hermitian spin--orbit-coupled fermions},
	author={Ren, Zejian and Liu, Dong and Zhao, Entong and He, Chengdong and Pak, Ka Kwan and Li, Jensen and Jo, Gyu-Boong},
	journal={Nature Physics},
	volume={18},
	number={4},
	pages={385--389},
	year={2022},
	publisher={Nature Publishing Group UK London}
}

@article{zhao2023two, 
year = {2025}, 
title = {{Two-dimensional non-Hermitian skin effect in an ultracold Fermi gas}}, 
author = {Zhao, Entong and Wang, Zhiyuan and He, Chengdong and Poon, Ting Fung Jeffrey and Pak, Ka Kwan and Liu, Yu-Jun and Ren, Peng and Liu, Xiong-Jun and Jo, Gyu-Boong}, 
journal = {Nature}, 
pages = {565--573}, 
number = {8046}, 
volume = {637}, 
}

@article{zhou2023non,
	title={Non-Abelian generalization of non-Hermitian quasicrystals: PT-symmetry breaking, localization, entanglement, and topological transitions},
	author={Zhou, Longwen},
	journal={Physical Review B},
	volume={108},
	number={1},
	pages={014202},
	year={2023},
	publisher={APS}
}

@article{hatano1996localization,
  title={Localization transitions in non-Hermitian quantum mechanics},
  author={Hatano, Naomichi and Nelson, David R},
  journal={Physical review letters},
  volume={77},
  number={3},
  pages={570},
  year={1996},
  publisher={APS}
}

@article{Song2019, 
  year     = {2019}, 
  title    = {Observation of nodal-line semimetal with ultracold fermions in an optical lattice}, 
  author   = {Song, Bo and He, Chengdong and Niu, Sen and Zhang, Long and Ren, Zejian and Liu, Xiong-Jun and Jo, Gyu-Boong}, 
  journal  = {Nature Physics}, 
  doi      = {10.1038/s41567-019-0564-y}, 
  eprint   = {1808.07428}, 
  url      = {arXiv.org}, 
  pages    = {911 -- 916}, 
  number   = {9}, 
  volume   = {15}, 
  note     = {16 pages including supplementary material, 4 figures}, 
  month    = {09}
}

@article{Wu2016, 
  year     = {2016}, 
  title    = {Realization of two-dimensional spin-orbit coupling for Bose-Einstein condensates}, 
  author   = {Wu, Zhan and Zhang, Long and Sun, Wei and Xu, Xiao-Tian and Wang, Bao-Zong and Ji, Si-Cong and Deng, Youjin and Chen, Shuai and Liu, Xiong-Jun and Pan, Jian-Wei}, 
  journal  = {Science}, 
  doi      = {10.1126/science.aaf6689}, 
  pages    = {83 -- 88}, 
  number   = {6308}, 
  volume   = {354},  
  note     = {27 pages, 5 figures}, 
  month    = {10}
}

@article{Liang2023, 
  year     = {2023}, 
  title    = {Realization of Qi-Wu-Zhang model in spin-orbit-coupled ultracold fermions}, 
  author   = {Liang, Ming-Cheng and Wei, Yu-Dong and Zhang, Long and Wang, Xu-Jie and Zhang, Han and Wang, Wen-Wei and Qi, Wei and Liu, Xiong-Jun and Zhang, Xibo}, 
  journal  = {Physical Review Research}, 
  doi      = {10.1103/physrevresearch.5.l012006}, 
  eprint   = {2109.08885}, 
  pages    = {L012006}, 
  number   = {1}, 
  volume   = {5}
}

@article{He2019, 
  year     = {2019}, 
  title    = {Recent progresses of ultracold two-electron atoms}, 
  author   = {He, Chengdong and Hajiyev, Elnur and Ren, Zejian and Song, Bo and Jo, Gyu-Boong}, 
  journal  = {Journal of Physics B: Atomic, Molecular and Optical Physics}, 
  doi      = {10.1088/1361-6455/ab153e}, 
  pages    = {102001}, 
  number   = {10}, 
  volume   = {52}, 
  month    = {04}
}

@article{Li2019, 
  year     = {2019}, 
  title    = {Observation of parity-time symmetry breaking transitions in a dissipative Floquet system of ultracold atoms}, 
  author   = {Li, Jiaming and Harter, Andrew K. and Liu, Ji and Melo, Leonardo de and Joglekar, Yogesh N. and Luo, Le}, 
  journal  = {Nature Communications}, 
  doi      = {10.1038/s41467-019-08596-1}, 
  pmid     = {30787299}, 
  eprint   = {1608.05061}, 
  pages    = {855}, 
  number   = {1}, 
  volume   = {10}, 
  month    = {12}
}

@article{Jiang2019r9s, 
  year     = {2019}, 
  title    = {Interplay of non-Hermitian skin effects and Anderson localization in nonreciprocal quasiperiodic lattices}, 
  author   = {Jiang, Hui and Lang, Li-Jun and Yang, Chao and Zhu, Shi-Liang and Chen, Shu}, 
  journal  = {Physical Review B}, 
  issn     = {2469-9950}, 
  doi      = {10.1103/physrevb.100.054301}, 
  pages    = {054301}, 
  number   = {5}, 
  volume   = {100}
}

@article{zhang2021observation,
  title={Observation of non-Hermitian topology with nonunitary dynamics of solid-state spins},
  author={Zhang, Wengang and Ouyang, Xiaolong and Huang, Xianzhi and Wang, Xin and Zhang, Huili and Yu, Yefei and Chang, Xiuying and Liu, Yanqing and Deng, Dong-Ling and Duan, L-M},
  journal={Physical Review Letters},
  volume={127},
  number={9},
  pages={090501},
  year={2021},
  publisher={APS}
}

@article{cao2023probing,
  title={Probing complex-energy topology via non-Hermitian absorption spectroscopy in a trapped ion simulator},
  author={Cao, M-M and Li, Kai and Zhao, W-D and Guo, W-X and Qi, B-X and Chang, X-Y and Zhou, Z-C and Xu, Yong and Duan, L-M},
  journal={Physical Review Letters},
  volume={130},
  number={16},
  pages={163001},
  year={2023},
  publisher={APS}
}

@article{zhang2025observation,
  title={Observation of a non-Hermitian supersonic mode on a trapped-ion quantum computer},
  author={Zhang, Yuxuan and Carrasquilla, Juan and Kim, Yong Baek},
  journal={Nature Communications},
  volume={16},
  number={1},
  pages={3286},
  year={2025},
  publisher={Nature Publishing Group UK London}
}

@article{yu2022experimental,
  title={Experimental unsupervised learning of non-Hermitian knotted phases with solid-state spins},
  author={Yu, Yefei and Yu, Li-Wei and Zhang, Wengang and Zhang, Huili and Ouyang, Xiaolong and Liu, Yanqing and Deng, Dong-Ling and Duan, L-M},
  journal={npj Quantum Information},
  volume={8},
  number={1},
  pages={116},
  year={2022},
  publisher={Nature Publishing Group UK London}
}

@article{chen2021quantum,
  title={Quantum jumps in the non-Hermitian dynamics of a superconducting qubit},
  author={Chen, Weijian and Abbasi, Maryam and Joglekar, Yogesh N and Murch, Kater W},
  journal={Physical Review Letters},
  volume={127},
  number={14},
  pages={140504},
  year={2021},
  publisher={APS}
}

@article{harter2016pt,
  title={PT-breaking threshold in spatially asymmetric Aubry-Andr{\'e} and Harper models: Hidden symmetry and topological states},
  author={Harter, Andrew K and Lee, Tony E and Joglekar, Yogesh N},
  journal={Physical Review A},
  volume={93},
  number={6},
  pages={062101},
  year={2016},
  publisher={APS}
}

@article{shnerb1998winding,
  title={Winding numbers, complex currents, and non-Hermitian localization},
  author={Shnerb, Nadav M and Nelson, David R},
  journal={Physical review letters},
  volume={80},
  number={23},
  pages={5172},
  year={1998},
  publisher={APS}
}

@article{yuce2014pt,
  title={PT symmetric Aubry--Andr{\'e} model},
  author={Yuce, C},
  journal={Physics Letters A},
  volume={378},
  number={30-31},
  pages={2024--2028},
  year={2014},
  publisher={Elsevier}
}

@article{liang2014pt,
  title={PT restoration via increased loss and gain in the PT-symmetric Aubry-Andr{\'e} model},
  author={Liang, Charles H and Scott, Derek D and Joglekar, Yogesh N},
  journal={Physical Review A},
  volume={89},
  number={3},
  pages={030102},
  year={2014},
  publisher={APS}
}

@article{longhi2019topological,
  title={Topological phase transition in non-Hermitian quasicrystals},
  author={Longhi, Stefano},
  journal={Physical review letters},
  volume={122},
  number={23},
  pages={237601},
  year={2019},
  publisher={APS}
}

@article{hang2015localization,
  title={Localization of light in a parity-time-symmetric quasi-periodic lattice},
  author={Hang, Chao and Kartashov, Yaroslav V and Huang, Guoxiang and Konotop, Vladimir V},
  journal={Optics letters},
  volume={40},
  number={12},
  pages={2758--2761},
  year={2015},
  publisher={Optical Society of America}
}

@article{longhi2014PT,
doi = {10.1088/1751-8113/47/16/165302},
url = {https://dx.doi.org/10.1088/1751-8113/47/16/165302},
year = {2014},
month = {apr},
publisher = {IOP Publishing},
volume = {47},
number = {16},
pages = {165302},
author = {Longhi, Stefano},
title = {PT-symmetric optical superlattices},
journal = {Journal of Physics A: Mathematical and Theoretical}
}

@article{El-Ganainy2018, 
  year     = {2018}, 
  title    = {Non-Hermitian physics and {PT} symmetry}, 
  author   = {El-Ganainy, Ramy and Makris, Konstantinos G and Khajavikhan, Mercedeh and Musslimani, Ziad H and Rotter, Stefan and Christodoulides, Demetrios N}, 
  journal  = {Nat.Phys}, 
  doi      = {10.1038/nphys4323}, 
  pages    = {11 -- 19}, 
  number   = {1}, 
  volume   = {14}, 
}

@article{Ashida2021, 
  year     = {2021}, 
  title    = {Non-Hermitian physics}, 
  author   = {Ashida, Yuto and Gong, Zongping and Ueda, Masahito}, 
  journal  = {Advances in Physics}, 
  issn     = {0001-8732}, 
  doi      = {10.1080/00018732.2021.1876991}, 
  eprint   = {2006.01837}, 
  url      = {arXiv.org}, 
  pages    = {249--435}, 
  number   = {3}, 
  volume   = {69}, 
  note     = {Review article commissioned by Advances in Physics - 191 pages, 49 figures, 8 tables}
}

@article{Patil2022, 
  year     = {2022}, 
  title    = {Measuring the knot of non-Hermitian degeneracies and non-commuting braids}, 
  author   = {Patil, Yogesh S. S. and Höller, Judith and Henry, Parker A. and Guria, Chitres and Zhang, Yiming and Jiang, Luyao and Kralj, Nenad and Read, Nicholas and Harris, Jack G. E.}, 
  journal  = {Nature}, 
  issn     = {0028-0836}, 
  doi      = {10.1038/s41586-022-04796-w}, 
  pmid     = {35831605}, 
  eprint   = {2112.00157}, 
  pages    = {271--275}, 
  number   = {7918}, 
  volume   = {607}
}

@article{Lee2009, 
  year     = {2009}, 
  title    = {Observation of an Exceptional Point in a Chaotic Optical Microcavity}, 
  author   = {Lee, Sang-Bum and Yang, Juhee and Moon, Songky and Lee, Soo-Young and Shim, Jeong-Bo and Kim, Sang Wook and Lee, Jai-Hyung and An, Kyungwon}, 
  journal  = {Physical Review Letters}, 
  issn     = {0031-9007}, 
  doi      = {10.1103/physrevlett.103.134101}, 
  pmid     = {19905515}, 
  eprint   = {0905.4478}, 
  pages    = {134101}, 
  number   = {13}, 
  volume   = {103}
}

@article{zhou2023exact,
  title={Exact new mobility edges between critical and localized states},
  author={Zhou, Xin-Chi and Wang, Yongjian and Poon, Ting-Fung Jeffrey and Zhou, Qi and Liu, Xiong-Jun},
  journal={Physical Review Letters},
  volume={131},
  number={17},
  pages={176401},
  year={2023},
  publisher={APS}
}

@article{zhou2025fundamental,
  title={The fundamental localization phases in quasiperiodic systems: A unified framework and exact results},
  author={Zhou, Xin-Chi and Yao, Bing-Chen and Wang, Yongjian and Wang, Yucheng and Wei, Yudong and Zhou, Qi and Liu, Xiong-Jun},
  journal={arXiv preprint arXiv:2503.24380},
  year={2025}
}

@article{wang2021many,
  title={Many-body critical phase: extended and nonthermal},
  author={Wang, Yucheng and Cheng, Chen and Liu, Xiong-Jun and Yu, Dapeng},
  journal={Physical Review Letters},
  volume={126},
  number={8},
  pages={080602},
  year={2021},
  publisher={APS}
}

@article{Zhao2022, 
  year    = {2022}, 
  title   = {Observing a topological phase transition with deep neural networks from experimental images of ultracold atoms.}, 
  author  = {Zhao, Entong and Mak, Ting Hin and He, Chengdong and Ren, Zejian and Pak, Ka Kwan and Liu, Yu-Jun and Jo, Gyu-Boong}, 
  journal = {Optics express}, 
  doi     = {10.1364/oe.473770}, 
  pmid    = {36258360}, 
  pages   = {37786--37794}, 
  number  = {21}, 
  volume  = {30}
}

@article{schreiber2015observation,
  title={Observation of many-body localization of interacting fermions in a quasirandom optical lattice},
  author={Schreiber, Michael and Hodgman, Sean S and Bordia, Pranjal and L{\"u}schen, Henrik P and Fischer, Mark H and Vosk, Ronen and Altman, Ehud and Schneider, Ulrich and Bloch, Immanuel},
  journal={Science},
  volume={349},
  number={6250},
  pages={842--845},
  year={2015},
  publisher={American Association for the Advancement of Science}
}

@article{li2022engineering,
  title={Engineering dissipative quasicrystals},
  author={Li, Tianyu and Zhang, Yong-Sheng and Yi, Wei},
  journal={Physical Review B},
  volume={105},
  number={12},
  pages={125111},
  year={2022},
  publisher={APS}
}

@article{zhang2017two,
  title={Two-leg Su-Schrieffer-Heeger chain with glide reflection symmetry},
  author={Zhang, Shao-Liang and Zhou, Qi},
  journal={Physical Review A},
  volume={95},
  number={6},
  pages={061601},
  year={2017},
  publisher={APS}
}

@article{lang2017nodal,
  title={Nodal Brillouin-zone boundary from folding a Chern insulator},
  author={Lang, Li-Jun and Zhang, Shao-Liang and Zhou, Qi},
  journal={Physical Review A},
  volume={95},
  number={5},
  pages={053615},
  year={2017},
  publisher={APS}
}

@article{zhou2025recovering,
  title={Recovering dark states by non-Hermiticity},
  author={Zhou, Qi},
  journal={AAPPS Bulletin},
  volume={35},
  number={1},
  pages={1--8},
  year={2025},
  publisher={Springer}
}

@article{molignini2023anomalous,
  title={Anomalous skin effects in disordered systems with a single non-Hermitian impurity},
  author={Molignini, Paolo and Arandes, Oscar and Bergholtz, Emil J},
  journal={Physical Review Research},
  volume={5},
  number={3},
  pages={033058},
  year={2023},
  publisher={APS}
}

@article{molignini2025stability,
  title={Stability of quasicrystalline ultracold fermions to dipolar interactions},
  author={Molignini, Paolo},
  journal={Physical Review Research},
  volume={7},
  number={3},
  pages={L032026},
  year={2025},
  publisher={APS}
}

@article{molignini2025stability2,
  title={Stability of dipolar bosons in a quasiperiodic potential},
  author={Molignini, Paolo and Chakrabarti, Barnali},
  journal={Physical Review Research},
  volume={7},
  number={2},
  pages={023237},
  year={2025},
  publisher={APS}
}

@article{zhou2022engineering,
  title={Engineering non-Hermitian skin effect with band topology in ultracold gases},
  author={Zhou, Lihong and Li, Haowei and Yi, Wei and Cui, Xiaoling},
  journal={Communications Physics},
  volume={5},
  number={1},
  pages={252},
  year={2022},
  publisher={Nature Publishing Group UK London}
}

@article{okuma2020topological,
  title={Topological origin of non-Hermitian skin effects},
  author={Okuma, Nobuyuki and Kawabata, Kohei and Shiozaki, Ken and Sato, Masatoshi},
  journal={Physical review letters},
  volume={124},
  number={8},
  pages={086801},
  year={2020},
  publisher={APS}
}

@article{zhang2020correspondence,
  title={Correspondence between winding numbers and skin modes in non-Hermitian systems},
  author={Zhang, Kai and Yang, Zhesen and Fang, Chen},
  journal={Physical Review Letters},
  volume={125},
  number={12},
  pages={126402},
  year={2020},
  publisher={APS}
}

\end{document}